%% file: main.tex
\documentclass[sigconf,natbib]{acmart}

\usepackage{algorithm}
\usepackage{multirow}
\usepackage{longtable}
\usepackage{graphicx}
\usepackage{adjustbox}
\usepackage{booktabs}
\usepackage{colortbl}
\usepackage{multirow}
\usepackage{rotating}
\usepackage{tabularx}

\usepackage[show]{chato-notes}

\DeclareMathOperator*{\argmax}{argmax} %

\usepackage[neverdecrease]{paralist}
\usepackage{enumitem} 
\newcommand{\uls}{\begin{itemize}[leftmargin=*]}
\newcommand{\ule}{\end{itemize}}
\newcommand{\ols}{\begin{enumerate}[leftmargin=*]}
\newcommand{\ole}{\end{enumerate}}
\newcommand{\li}{\item}
\newcommand{\nores}{\cellcolor{lightgray}}

\newcommand{\para}[1]{\paragraph{\textnormal{\textbf{#1}}}} 

\copyrightyear{2023}
\acmYear{2023}
\setcopyright{acmlicensed}\acmConference[SIGIR '23]{Proceedings of the 46th
International ACM SIGIR Conference on Research and Development in
Information Retrieval}{July 23--27, 2023}{Taipei, Taiwan}
\acmBooktitle{Proceedings of the 46th International ACM SIGIR Conference on
Research and Development in Information Retrieval (SIGIR '23), July 23--27,
2023, Taipei, Taiwan}
\acmPrice{15.00}
\acmDOI{10.1145/3539618.3591982}
\acmISBN{978-1-4503-9408-6/23/07}

\begin{document}

\title{Explain like I am BM25: Interpreting a Dense Model's Ranked-List with a Sparse Approximation}

\author{Michael Llordes}
\email{micllrs@gmail.com}
\affiliation{%
  \institution{University of Glasgow}
  \city{Glasgow}
  \country{United Kingdom}
}

\author{Debasis Ganguly}
\email{debasis.ganguly@glasgow.ac.uk}
\affiliation{%
  \institution{University of Glasgow}
  \city{Glasgow}
  \country{United Kingdom}
  }

\author{Sumit Bhatia}
\email{sumit.bhatia@adobe.com}
\affiliation{%
  \institution{Media and Data Science Research Lab, Adobe}
  \city{Noida}
  \country{India}
  }

\author{Chirag Agarwal}
\email{chiragagarwall12@gmail.com}
\affiliation{%
  \institution{Media and Data Science Research Lab, Adobe}
 \city{Noida}
  \country{India}
}

\renewcommand{\shortauthors}{Llordes, et al.}

\begin{abstract}

\input{files/abstract_ML.tex}

\end{abstract}

\ccsdesc[500]{Information systems~Information retrieval}
\ccsdesc[500]{Information systems~Content analysis and feature selection}
\ccsdesc[500]{Information systems~Retrieval models and ranking}

\keywords{Interpretability, Explainability, Neural Ranking Models}

\maketitle

\input{files/introduction}

\input{files/problem}

\input{files/experiments}

\input{files/discussion.tex}

\input{files/conclusion.tex}

\bibliographystyle{ACM-Reference-Format}
\bibliography{ref.bib}

\end{document}

%% file: files/abstract_ML.tex
Neural retrieval models (NRMs) have been shown to outperform their statistical counterparts owing to their ability to capture semantic meaning via dense document representations. These models, however, suffer from poor interpretability as they do not rely on explicit term matching. As a form of local per-query explanations, we introduce the notion of \textit{equivalent} queries that are generated by maximizing the similarity between the NRM’s results and the result set of a sparse retrieval system with the equivalent query. We then compare this approach with existing methods such as RM3-based query expansion and contrast differences in retrieval effectiveness and in the terms generated by each approach.

%% file: files/introduction.tex
\section{Introduction}
\label{Sec:intro}

Neural retrieval models (NRM) have gained prominence, achieving state-of-the-art results on various document and passage ranking tasks~\cite{khattab2020colbert,monot5,ance,deepct,nogueira2019doc2query}. NRMs are capable of modeling the semantic similarity between the query and document representations for ranking, leading to their outperforming over traditional sparse retrieval methods (e.g., BM25 \cite{Okapi}, LM \cite{lmdir} etc.), which rely explicitly on term matching. However, despite their success, NRMs suffer from poor \textit{interpretability} of their results~\cite{explainable-ir-survey}. With the increasing deployment of more complex NRMs, it is essential to explain the retrieval decisions of a ``black-box'' complex model to its end-users, thereby increasing their trust in the model~\cite{qu2021deep}.

Given a query, while it is straightforward to understand why a document is retrieved by a sparse retrieval model as the relevance score depends on the explicit presence of the query (or expansion) terms, the results produced by NRMs are hard to interpret as these models rely on the closeness of query and document representations in the embedding space. This inherent opaqueness of NRMs makes them potentially non-trustworthy to end-users, especially in critical domains such as healthcare, finance, and law~\citep{10.1145/3209978.3210193, explainable-ir-survey}. Approaches to explaining NRMs include providing additional information such as informative snippets~\citep{explainable-elements} and key noun phrases representing query aspects~\cite{docrelevance2search}, visualizing regions in the embedding space that affect the model output~\citep{gradcam} and personalized explanations based on user characteristics~\citep{verberne2018explainable}. Furthermore, it has been found that explanations generated by popular interpretabilty models such as LIME~\cite{lime} and SHAP~\cite{shap,deepshap} vary widely raising concern about the robustness and utility of the explanations produced by directly applying these interpretability methods to NRMs~\citep{deepshap}.
\newline
\\ \noindent 
\textbf{Our Contributions:}
We propose an intuitive means of explaining the output of an NRM by introducing the notion of an \textit{equivalent query}, which we define as follows. Given a query $Q$, the equivalent query $Q^+$ is the query which when executed on a sparse retrieval model ideally produces the same ranked list as produced by an NRM with $Q$, the original query. We posit that the equivalent query offers a minimalist and intuitive explanation of the ``thought process'' of an NRM. Since a sparse IR model relies on explicit term matching, a query that produces the same (or close enough) results to that of a complex NRM reveals the semantic concepts considered implicitly by the NRM, and thus potentially helps to interpret its behavior. %

As an illustrative example, consider the query \emph{what is the most popular food in switzerland} from the TREC-DL'19 topic set. The equivalent queries produced by our proposed method for Mono-T5~\cite{monot5} (an NRM) is `\textit{dish food includ serv switzerland vacherin}' whereas for DCT~\cite{deepct} (another NRM) this equivalent query is `\emph{appenzel food german meat neighbor popular}'. The interesting point is that `appenzeler' and `vacherin' are two different varieties of Swiss cheese; the two equivalent queries produced by our method unravel the different concepts and terms used by these two models for retrieving the respective top documents.

While the notion of an `equivalent query' may seem somewhat similar to adding terms to the original query via pseudo-relevance feedback (PRF), there are two major differences. First, expansion terms in PRF are simply the informative terms occurring in the top-documents, whereas the process of generating equivalent queries explicitly seeks to make the output of a sparse model locally similar to that of a target NRM. Second, unlike PRF expansion, terms from the original query may also be absent in the equivalent query.

Constructing the equivalent queries as explanations for a particular NRM is non-trivial as finding such an optimal query
is an optimal subset-selection problem, which is NP-complete (Section~\ref{sec:problem}).
We
adopt a discrete state-space exploration method to find solution states (expanded queries) which when executed on a sparse index maximises the overlap of the top-documents retrieved with the NRM for which explanations are sought. %
Our experiments on the MS-MARCO~\cite{nguyen2016ms} reveal that the generated equivalent queries, on an average,
achieve a fidelity score (RBO \cite{rbo_paper}) of up to 0.5194, and these equivalent queries when executed on BM25 lead up to obtaining 96\% of the nDCG values of the target NRM.

%% file: files/problem.tex
\section{Proposed Methodology}
\label{sec:problem}

\para{Problem Formulation}
Let $\theta$ be a neural ranking model (NRM) that, given a query $Q$, retrieves an ordered set of documents $L_k(Q;\theta)$. Each document $D \in L_k(Q;\theta)$ is sorted in the decreasing order of the parameterized scores  $\theta(Q, D)$ that measure the relevance of $D$ to $Q$. The dense model $\theta$ can either re-rank an initial list produced by a sparse retriever (e.g. BM25)~\cite{khattab2020colbert}, or retrieve the list directly from the underlying corpus in an end-to-end manner~\cite{ance}.
Recall that our goal is to make use of a sparse retrieval model, which we denote as $\phi$ (e.g., BM25 \cite{Okapi} or LM \cite{lmdir}) to approximate the retrieval output of an NRM $\theta$. Note that this idea is analogous to the existing work on developing local surrogate models to explain the behavior of complex black-box models~\cite{lime,shap,L2X}.

The sparse IR model, when fed with the equivalent query to approximate the results of an NRM, offers a surrogate model for explaining the behavior of the target NRM. This is because the sparse model i) leverages discrete terms instead of embeddings, and ii) the scoring function $\phi(Q, D)$ is a closed form expression of several basic components, such as the term frequencies and IDFs of the matching terms, the length of $D$, etc.~\cite{DBLP:conf/sigir/SenGVJ20,axiomatic}.

There is one subtle difference of explaining document ranking from the per-instance based local explanation models for classifiers (e.g., LIME~\cite{lime} or SHAP~\cite{shap}) that explain the output by inducing weights over input features reflecting their relative influence on the predicted outcome. In the context of document ranking, similar ideas have been explored to measure the impact of individual terms on relative changes in the document scores \cite{verma2019lirme}.

In our setting, the key difference is that the \textit{locality for approximation} to estimate term influence is not restricted to individual instances of query-document pairs. Instead, the approximator works at the level of a query and the top-$k$ set of documents retrieved with a deep neural model.  
Due to this difference in the granularity of locality, the explanations generated do not correspond to the influence weights of words from individual documents, but rather they correspond to the words from the top-retrieved set. 
Specifically, we seek to find those terms which when added to the original query $Q$ will effectively bridge the vocabulary gap of the sparse model $\phi$ and make its output similar to that of the NRM ($\theta$) for $Q$.
Formally speaking, the \textit{equivalent query} $Q^+$ is the one that satisfies the following objective:
\begin{equation}
\argmax_{Q^+ \subset V(L_k(Q; \theta))} \omega(L_k(Q^+; \phi), L_k(Q; \theta)) \label{eq:objective}, 
\end{equation}
where $V(L_k(Q; \theta))$ represents the vocabulary of the top-$k$ documents retrieved with the query $Q$ using the model $\theta$, and $\omega$ is a similarity measure, e.g., the set-based Jaccard metric or the rank-based RBO metric~\cite{rbopaper} between the two ranked lists of documents (see also \cite{DBLP:conf/cikm/Sen0GVR22} which explored this idea of overlap between document lists retrieved with two different queries for measuring trustworthiness of NRMs). The output query, $Q^+$, thus obtained, can be interpreted as the set of terms, or \textit{concepts}, that the black-box NRM takes into account in its computation of the top-$k$ list. While in reality, $\theta$ works in the embedded (continuous) space, a discrete realisation of this concept set via this approximation helps gain insights into the behavior of $\theta$, which would be useful to end-users and model practitioners.

\para{Discrete State-Space Optimisation} Observe that Equation~\ref{eq:objective} is an optimal subset selection problem, which is NP-complete. A practical approach is to employ a standard discrete state-space exploration method, such as the Best First Search (BFS) exploration~\cite{russell2016artificial} that involves traversing a state-space transition tree by dynamically selecting a depth-first or breadth-first strategy via a heuristic. We now describe the state-space, the actions and the heuristics employed in the BFS exploration for our task.

\para{States and Evaluation Function} In the context of our problem, a state refers to a query $Q^+$ which is executed by the sparse model $\phi$ to retrieve a top-$k$ list. The goal state for this approximation problem is to find an optimal query $Q^*$ which retrieves a top-$k$ set identical to the black-box model $\theta$. Thus, the evaluation function for a state, which measures how close a state is to the goal state, is the overlap measure shown in Equation \ref{eq:objective}.

\para{Actions} We consider two types of actions - one that involves adding terms to an existing query of a state to generate a new query state, and the other that involves removing a term to generate a new state. An important decision to be made is to determine the set of candidate terms to be added to an existing state to create a new state. Although, in principle, one can consider the entire set of vocabulary terms, such an approach would lead to a substantially large branching factor for the tree-based exploration. A large volume of work on pseudo-relevance feedback (PRF) in IR has shown that most terms that are semantically related to the information need mostly occur within the top-$k$ set of documents retrieved \cite{DBLP:conf/cikm/RoyGMJ16,DBLP:conf/sigir/Montazeralghaem20,DBLP:conf/cikm/ChakrabortyGC20}. Thus, it is reasonable to restrict the set of candidate terms to the vocabulary of this set, denoted as $V(L_k(Q; \theta))$ in Equation \ref{eq:objective}.

To define the first type of transition, i.e., the one that involves adding a term $t$ to transition from $Q_i$ to $Q_{i+1} = Q_i \cup \{t\}$ ($i$ being the depth of the BFS exploration tree), is given by
the normalized probability of the RM3 \cite{Lavrenko_RLM2001:RBL:383952.383972} weights. In other words, the higher the RM3 weight of a term $t \in V(L_k(Q_{i+1}; \theta))$, the higher is the likelihood of exploring along the branch $Q_{i+1}$. Similarly, for generating a new query (state) with one term removed from the current state, we set the probability of removing a term as \textit{inversely proportional} to its tf-idf weight with the rationale of retaining the informative terms within a query.

\para{Exploration Heuristic} For the BFS exploration, we start exploring from the root state - the empty query $\emptyset$. Exploring a state involves generating $b$ child states by choosing one of the two actions randomly: add or remove ($b$ denotes the maximum branching factor parameter). A child state is only added to the tree if it has not been generated before. throughout the tree exploration phase, we keep track of the child states, i.e., queries generated thus far.

Given a set of current unexplored nodes $S \in \mathcal{U}$, the next state considered for exploration (action of adding or removing terms to generate newer queries) is the state (say $S^* \in \mathcal{U}$) with the best value of the evaluation function, i.e., $\argmax_{S \in \mathcal{U}} \omega(L_k(S; \phi), L_k(Q; \theta))$. An advantage of the best-first exploration is that it is able to continue exploring along a promising direction at greater depths, or is able to back-track to expand yet unexplored states at lower depths. During the execution of the algorithm, we keep track of the best state discovered and output it at the end of the exploration. The exploration itself is limited by the maximum number of depths, which we set to $10$ in our experiments. The termination condition for the algorithm is given by this maximum number of depths, or when there is no state left to explore.%

%% file: files/experiments.tex
\section{Experiment Setup}
\label{sec:experiments}

\para{Research Questions and Dataset} The objective of our experiments is to see how effectively can we approximate the top-retrieved documents of an NRM $\theta$ by a sparse model $\phi$, i.e.,
\uls
\li \textbf{RQ1}: How well our proposed method of BFS-based tree exploration approximates several black-box models (e.g., ColBERT \cite{khattab2020colbert}, ANCE \cite{ance} etc.) with different modes of operation (sparse with reranking, or end-to-end dense with approximate search)?
\ule
Since the output of a model-aware local approximation is a set of additional terms, which are supposed to be those on which the target NRM puts emphasis, 
the next question to investigate is:
\uls
\li \textbf{RQ2}: Can these additional terms, on top of help interpreting $\theta$, can also help to improve the IR effectiveness of sparse models?
\ule
As the dataset for our experiments, we use the MS-MARCO passage ranking collection \cite{nguyen2016ms} and the TREC DL 2019 topic set \cite{craswell2021trec}.

\para{Baselines} Since we propose to use BFS to solve the discrete state-space optimisation of optimal subset selection, we compare it with other computationally less intensive approaches, such as the greedy search. In the greedy exploration of the state-space, we generate $b$ branches similar to the BFS method; however, we keep on exploring only along the best branch every time without saving the other branches for back-tracking purposes. Also, similar to the BFS method, in our greedy baseline we restrict the exploration to a maximum number of states and output the best state discovered during the exploration as the optimal solution. This baseline uses the same overlap-based state evaluation as used by BFS.

In addition to the \textit{model-aware} greedy baseline, we also employ a \textit{model-agnostic} baseline method, namely RM3-based query expansion \cite{Lavrenko_RLM2001:RBL:383952.383972}, to find out how much of an overlap can a model-agnostic method such as RM3 on a sparse index achieve with the top-retrieved set obtained by an NRM. Note that the output of this method cannot be used as a model-specific explanation, and merely serves as a reference point for the overlap comparison.

\para{Parameter Details} To solve the optimisation of Equation \ref{eq:objective}, for the greedy approach we set the maximum number of unique states visited to $1000$, whereas for BFS, we set the tree depth to $10$. The branching factor $b$ of the BFS exploration was set to $30$.
We set the parameter $k$ of Equation \ref{eq:objective}, which controls the locality of the explanations, to a value of $10$ in all our experiments.
A small value of $k=10$ ensures that the objective is to approximate only the first search result page of a black-box model~\cite{serpsize}.

\para{Evaluation Metrics} As a measure of how closely the sparse approximation fits a black-box model, i.e., as a fidelity measure, we report the RBO and Jaccard based overlaps between the top-10 documents retrieved with the original query by $\theta$ (the black-box) and those retrieved with the expanded query by BM25 ($\phi$, the approximator). As IR evaluation measures, we report the MAP (relevance of at least 2 as per the TREC DL guidelines \cite{craswell2021trec}) and the nDCG values for the top-$10$ results. A value of $k=10$ was used to optimise the sparse approximation objective (see Equation \ref{eq:objective}).

\para{Black-box and the Sparse Approximator Models}
As the sparse model for approximation we employ BM25, i.e., $\phi=\{\text{BM25}\}$ with the standard settings of $k_{\text{BM25}}=1.2$ and $b_{\text{BM25}}=0.75$. As a concrete realisation of the state evaluation function $\omega$ of Equation \ref{eq:objective}, we use RBO, which is a rank-based overlap metric \cite{rbopaper}. In our results, we report both Jaccard and RBO measures computed between the top-retrieved sets of $\theta$ and $\phi$. %
As the black-box model $\theta$, we employ a number of neural models operating in both the `sparse + reranking' and the dense end-to-end modes \cite{DBLP:conf/sigir/LiWZMMLZ22}. Specifically, we used ANCE~\cite{ance}, ColBERT (CBERT)~\cite{khattab2020colbert}, and MonoT5~\cite{monot5} as representative dense end-to-end models. Additionally, we employed two sparse-reranking based models, a DeepCT~\cite{deepct} augmented index with ColBERT (DCT), and ColBERT followed by a BERT-based query expansion (QE$_{BERT}$).
Our implementation is available at \url{https://github.com/micllordes/eliBM25}.

\section{Results} 
\label{sec:results}

\input{tabledefs/newRBOmainres}

\para{Main observations}
Tables~\ref{irtable} and \ref{fidelitytable} present the results of our experiments on approximating neural models with the sparse approximator - BM25; following are the interesting observations.
\textit{First}, in relation to RQ1, we observe that our proposed BFS-based optimisation consistently outperforms the baseline greedy approach by a large margin in terms of both the fidelity and the approximation, and also in terms of the quality of the search list retrieved with the model-aware expanded query $Q^+$ on BM25 (Table 2).

\textit{Second}, in relation to RQ2, we observe that the MAP and the nDCG values obtained via approximation are close to the values obtained with the black-box models themselves, e.g., compare the MAP value of $0.2064$ obtained with the sparse approximation to that of $0.2182$ obtained with CBERT (Table 1). Another important observation in relation to RQ2 is that the BFS-based model-specific approximation yields queries that are qualitatively better than the ones generated by an unsupervised relevance feedback model, such as RM3. This can be seen by comparing the MAP and the nDCG values of RM3 with the ones obtained by BFS approximation.
\textit{Third}, we see that high fidelity scores correlate well with the downstream task (retrieval) performance obtained via the approximation, which also indicates that the enriched query acts as a meaningful explanation of the model-specific influence of term weights.

\begin{figure}[t]
    \centering
    \includegraphics[width=\columnwidth]{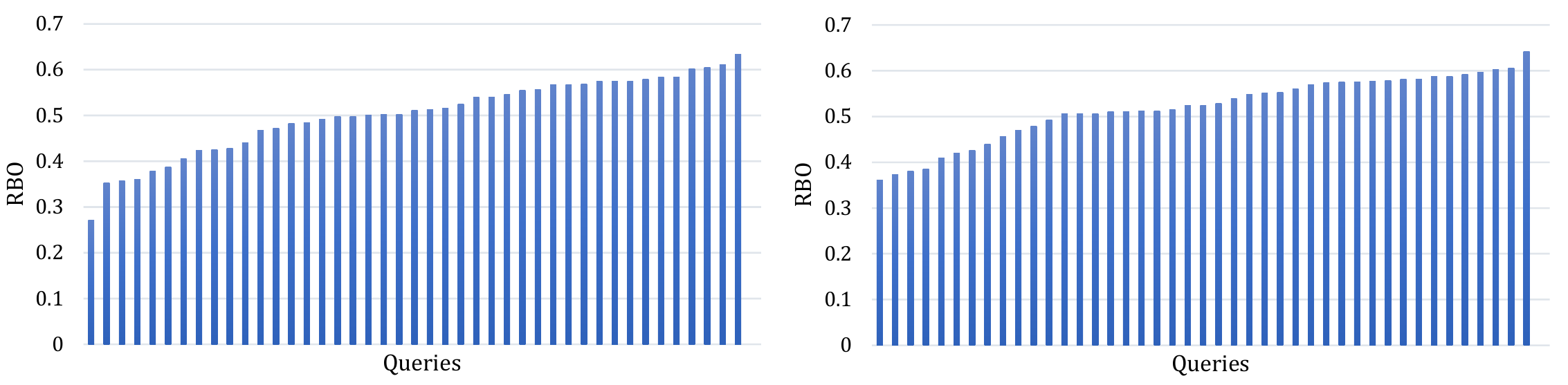}
    \caption{\small Fidelity (RBO overlap of top-10 documents) of the sparse approximation of the best performing models among the reranking ones - CBERT+BERTQE (\textit{left}), and the best of the end-to-end ones - MonoT5 (\textit{right}). The queries are sorted by the RBO values.}
    \label{fig:perq-jaccard}
\end{figure}

\para{Per-query statistics}
Figure \ref{fig:perq-jaccard} shows that a majority of the queries exhibit relatively high RBO values indicating that most of the queries are good candidates for model-specific explanations via our proposed discrete state-space optimisation. We believe that queries with small fidelity scores are a result of the limitation of the size of the vocabulary $V(L_k(Q; \theta))$ as described in Section \ref{sec:problem}.

%% file: tabledefs/newRBOmainres.tex
\begin{table}[t]
\centering
\caption{
\small
Results of sparse approximation of NRMs in terms of IR effectiveness.
The best results across each method (Greedy and BFS) are bold-faced and the ones across each column are under-lined.
The gray cells corresponding to the sparse retrieval rows (BM25 and RM3) mean that these models, being sparse ones themselves, are not approximated by the baseline (greedy) or the proposed approach (BFS).
All nDCG values obtained with the BFS approach show statistically significant differences compared to the baselines.
\label{irtable}}
\begin{adjustbox}{width=0.85\columnwidth}
\begin{tabular}{@{}ll cc cc cc@{}} 
\toprule
& & \multicolumn{2}{c}{} & \multicolumn{4}{c}{Retrieval quality with BM25$(Q^+)$} \\
\cmidrule{5-8}
Model & & \multicolumn{2}{c}{} & \multicolumn{2}{c}{Greedy (Baseline)} & \multicolumn{2}{c}{BFS (Ours)} \\
\cmidrule(l){5-6} \cmidrule(l){7-8}
Type & IR Model & MAP & nDCG & MAP & nDCG & MAP & nDCG \\ 
\midrule
\multirow{2}{*}{Sparse}

& BM25 & 0.1067 & 0.4601 & \nores & \nores & \nores & \nores \\
& RM3 & 0.1411 & 0.4931 & \nores & \nores & \nores & \nores \\ 
\midrule
\multirow{2}{*}{Rerank}
& DCT & 0.2192 & 0.7006 & 0.1544 &  0.5082 & \textbf{0.2050} & \textbf{0.6540}\\ 
& QE$_{BERT}$ & 0.2199 & 0.7015 & 0.1414 & 0.4848 & \textbf{\underline{0.2065}} & \textbf{\underline{0.6760}} \\ 
\midrule
\multirow{3}{*}{E2E}
& ANCE & 0.1836 & 0.6537 & 0.1454 & 0.5511 & \textbf{0.1723} & \textbf{0.6049} \\
& CBERT & 0.2182 & 0.6934 & 0.1506 & 0.5057 & \textbf{0.2064} & \textbf{0.6474} \\ 
& MonoT5 & 0.2184 & 0.7300 & 0.1470 & 0.5355 & \textbf{0.1920} & \textbf{0.6623} \\
\bottomrule
\end{tabular}
\end{adjustbox}
\end{table}

\begin{table}[t]
\centering
\caption{
\small
Evaluating sparse approximation of dense black-box models in terms of fidelity (overlap).
The naming and bold-face/underline conventions are the same as that of Table \ref{irtable}. All fidelity scores from our BFS approach show statistically significant differences from their corresponding BM25, RM3 and Greedy baselines.
\label{fidelitytable}}
\begin{adjustbox}{width=1\columnwidth}
\begin{tabular}{@{}ll cc cccc cccc@{}} 
\toprule
& & \multicolumn{8}{c}{Fidelity (Overlap) Measures}\\
\cmidrule(l){3-10}
Model & & \multicolumn{2}{c}{BM25$(Q)$} & \multicolumn{2}{c}{RM3} & \multicolumn{2}{c}{Greedy (Baseline)} & \multicolumn{2}{c}{BFS (Ours)} \\
\cmidrule(l){3-4} \cmidrule(l){5-6} \cmidrule(l){7-8} \cmidrule(l){9-10}
Type & IR Model & Jac & RBO & Jac & RBO & Jac & RBO & Jac & RBO\\ 
\midrule

Rerank
& DCT & 0.1883 & 0.1173 & 0.1694 & 0.0956 & 0.3140 & 0.2207 & \textbf{{0.5194}} & \textbf{0.4946} \\ 
(CBERT) & QE$_{BERT}$ & 0.1621 & 0.1144 & 0.1420 &  0.0916 & 0.2956 & 0.2213  & \textbf{0.5111} & \textbf{{0.5015}} \\ 
\midrule
\multirow{3}{*}{E2E}
& ANCE & 0.1501 & 0.0952 & 0.1256 & 0.0753 & 0.3106 & 0.2239 & \textbf{0.4993} & \textbf{0.4969} \\
& CBERT & 0.1641 & 0.1155 & 0.1442 & 0.0907 & 0.2974 & 0.2230 & \textbf{0.5046} & \textbf{0.4888} \\ 
& MonoT5 & 0.1835 & 0.1120 & 0.1715 & 0.0970 & 0.3041 & 0.2224 & \textbf{\underline{0.5327}} & \textbf{\underline{0.5194}} \\
\bottomrule
\end{tabular}
\end{adjustbox}
\end{table}

%% file: files/discussion.tex
\para{Example queries generated}
As an additional analysis, we now present in Table \ref{tab:posexample} some sample queries generated by the sparse approximation method and compare those with the RM3-based expanded ones. 
The first set of rows in Table \ref{tab:posexample} presents a situation, where the sparse equivalent query $Q^+$ is a subset of the RM3 terms. This particular example is notable as the BFS-generated equivalent query correctly extracts terms related to the true information need of the original query from the RLM expanded one, i.e., eliminating terms related to the Commonwealth of British origin, and retaining the ones related to the Commonwealth with that of Soviet origin. 

The second group of rows in Table \ref{tab:posexample}, shows an instance of BFS-generated query which achieved a Jaccard score of 1, i.e., BM25 when executed with the equivalent query $Q^+$ yields the identical top-10 documents as MonoT5 does with the original query. This shows that the equivalent query, in this case, represents an effective explanation of the term semantics with which MonoT5 operates for the original query.

\input{tabledefs/exampleres}

%% file: tabledefs/exampleres.tex
\begin{table}[t]
\centering
\caption{\small Sample equivalent queries $Q^+$ generated by optimising Equation \ref{eq:objective} (BFS-Gen). The BFS-Gen queries yield substantially better retrieval effectiveness than the original queries, or the RM3-expanded ones (RM3-Exp).
\label{tab:posexample}
}
\begin{adjustbox}{width=.9\columnwidth}
\begin{tabularx}{1.25\columnwidth}{@{}l@{~~}Xc@{~~}cc@{}}
\toprule
Query & Example Queries & \multicolumn{2}{c}{Fidelity} & \\ 
\cmidrule(l){3-4}
Type & (stemmed words) & Jac & RBO & nDCG \\ 
\midrule
Original & Who formed the commonwealth of independent states & 0.1764 & 0.3480 & 0.3695 \\
RM3-Exp & british commonwealth countri form independ nation republ soviet state union & 0.2500 & 0.3565 & 0.3034 \\
BFS-Gen & commonwealth soviet union state form & 0.5385 & 0.3934 & 0.6401 \\ 

\midrule
Original & what is durable medical equipment consist of & 0.4286 & 0.1350 & 0.5739 \\
RM3-Exp & benefit consist dme durabl equip ill item medic patient therapeut & 0.6670 & 0.2039 & 0.8533 \\
BFS-Gen & medic consist equip item patient & 1.0000 & 0.2126 & 0.8807 \\
\bottomrule
\end{tabularx}
\end{adjustbox}
\end{table}

%% file: files/conclusion.tex
\section{Conclusions and Future Directions}
We considered the task of explaining the results of an NRM and introduced the notion of an equivalent query -- one that when executed on a sparse retrieval model can lead to similar results as a complex NRM. We formulated the problem as an optimal subset selection problem (NP complete) and proposed a BFS-based state-space exploration method to approximate equivalent queries. Our empirical results on MS-MARCO benchmark show that the equivalent queries produced by our solution can approximate the top-$k$ results of NRMs such as  ColBERT and ANCE (Section~\ref{sec:results}), and the resulting queries unravel how the NRMs interpreted the user queries. Further, the equivalent queries when executed on a BM25 ranker achieved retrieval performance close to the complex NRMs. 

Our proposed framework of equivalent queries offers a simple and intuitive interpretation of complex black-box retrieval models. One limitation of our solution is high latency due to the exploration of the state space (average latency of $\approx 6$ seconds). Our future work will focus on exploring different subset selection methods to reduce the number of retrieval operations on the index needed by the search; exploring reinforcement learning for improving the latency at run-time by learning optimal state transitions. Further, another interesting direction is to use the equivalent queries for extracting better snippets from the retrieved documents.